# Title: Influence of sidestepping expertise and core stability on knee joint loading during change of direction

Authors:

Youri Duchene[ab]*, Gérome C. Gauchard[ab], and Guillaume Mornieux[ab]

[a]*Université de Lorraine, DevAH, F-54000 Nancy, France*

[b]*Université de Lorraine, Faculty of Sport Sciences, Nancy, France*

Author for correspondence:

Youri Duchene

EA 3450 Développement, Adaptation et Handicap (DevAH), Faculté de Médecine

9 avenue de la Forêt de Haye CS 50184

54505 VANDOEUVRE LÈS NANCY

Tel.: +33 (0)3 83 68 37 45

Email: youri.duchene@univ-lorraine.fr

Authors ORCiDs and media:

Youri Duchene. ORCiD: https://orcid.org/0000-0002-2774-2333; Twitter: @DucheneYouri.

Gérome C. Gauchard. ORCiD: https://orcid.org/0000-0002-2489-9343.

Guillaume Mornieux. ORCiD: https://orcid.org/0000-0001-5756-9642.

**Acknowledgments**

The authors would like to thank all the participants of this study and Kristin L. Sainani for the statistical help.

**Ethical approval**

All procedures were approved by the ethics committee Sud Mediterranee III (approval reference 2018.07.03 bis) and conformed to requirements of the Declaration of Helsinki.

**Abstract:**

The aims of this study were twofold: firstly, to compare core stability and knee joint loading between sidestepping experts and nonexperts; secondly, to determine core predictors of knee joint loading. Thirteen handball male players (experts) and fourteen karatekas (nonexperts) performed six unanticipated 45° sidestepping manoeuvers, while trunk and pelvis 3D kinematics as well as ground reaction forces were measured, and peak knee abduction moment (PKAM) was determined. Student t-tests enabled a comparison of both groups and a linear mixed model approach was used to identify PKAM predictors. Sidestepping experts demonstrated significantly lower pelvis rotation towards the new movement direction at the initial contact than nonexperts (4.9° vs. 10.8°) and higher PKAM (0.539 vs. 0.321 Nm/kg-bwt). Trunk medial lean, trunk axial rotation and pelvis anterior tilt at the initial contact predicted PKAM, while trunk axial rotation, pelvis medial lean and posterior ground reaction force predicted PKAM during the weight acceptance phase. Despite higher PKAM, handball players might not be at a higher risk of anterior cruciate ligament injury as the knee joint loading remained at a relatively low level during this sidestepping task. Core stability, in its three dimensions, is a key determinant of knee joint loading.

**Introduction**

Anterior cruciate ligament (ACL) rupture is a major injury occurring mostly during noncontact changes of direction or landings (Boden et al., 2000). Different biomechanical risk factors at the knee level (Alentorn-Geli et al., 2009) but also related to the trunk position and control (Hughes, 2014), have been identified in the literature.

Indeed, an elevated knee joint loading possibly stems from higher lateral trunk motion (Hewett & Myer, 2011; Jamison et al., 2012) and is also associated with trunk rotation (Critchley et al., 2020; Dempsey et al., 2007; Dempsey et al., 2012). Moreover, Zazulak et al. (2007) highlighted a potential connection between core stability and ACL injury risk in female athletes. Recent studies provided an analysis of trunk neuromuscular control to further understand trunk motion during landings (Haddas et al., 2016; Lessi et al., 2017) and changes of direction (Jamison et al., 2013; Oliveira et al., 2013a; Oliveira et al., 2013b). Accordingly, future research on trunk control during changes of direction and its consequences on knee joint injury risk should not only determine trunk kinematics but also analyze the related neuromuscular control, for instance based on trunk muscle co-contractions (Jamison et al., 2013).

ACL injury risk is dependent on different factors, including gender or the type of sport (Beynnon et al., 2014; Renstrom et al., 2008). Moreover, an influence of the skill level on knee joint loading could be expected as high-level athletes are supposed to perform sidestepping changes of direction more efficiently (Fujii et al., 2014; Lockie et al., 2014). For instance, Sigward and Powers (2006b) reported higher knee joint moments for elite athletes, whereas greater knee abduction angle was found for novices but without any abduction moment difference in another study (Kipp et al., 2013). Likewise, the influence of athletic background on knee joint biomechanics during various tasks associated with high knee joint loading remains controversial in the literature (Cowley et al., 2006; Herrington, 2011; Orishimo et al., 2014).

In addition, it has been demonstrated that athletic background has an impact on trunk control during unstable sitting tasks and trunk perturbations (Barbado et al., 2016; Glofcheskie & Brown, 2017). Differences between high-level sport athletes, i.e. judokas vs. kayakers (Barbado et al., 2016) and runners vs. golfers (Glofcheskie & Brown, 2017), were only noticed during specific tasks that were similar to those realized during their practice. Therefore, the comparison of task experts and nonexperts might give a more accurate insight of an adequate trunk control while performing changes of direction.

More generally, researchers seeking to further understand the knee joint injury risk and therefore predict peak knee abduction moment (PKAM) have recently provided models putting different trunk and lower limb variables together (Frank et al., 2013; Jones et al., 2015; Kristianslund et al., 2014; Staynor et al., 2020; Weir et al., 2019). While trunk lateral lean and trunk rotation were predictive variables for PKAM, little is known about the influence of the pelvis, despite its central role within the core stability. Also, while Jamison et al. (2013) underlined the effect of erector spinae muscles on knee joint loading and trunk flexion, the whole comprehension of the influence of core neuromuscular control (including glutei muscles, erector spinae muscles and abdominal muscles) on trunk and pelvis 3D motion during sidestepping maneuvers is lacking. Such a modeling approach, using data from a wide range of level of expertise in changes of direction, might provide a better understanding of the effect of the core kinematics and neuromuscular control on injury predictive variables during sidestepping tasks and help improve prevention programs to decrease the knee injury risk.

Thus, the purpose of this study was i) to determine the difference in core stability and knee joint loading between task experts and nonexperts during sidestepping maneuvers, ii) to understand core parameters predicting peak knee abduction moment, and iii) to determine neuromuscular parameters controlling core kinematics. We hypothesized that experts will have their trunk more orientated in the targeted direction, a higher neuromuscular co-contraction ratio towards this direction, and higher PKAM than nonexperts. Moreover, we suggested that PKAM will be predicted by trunk lateral lean and trunk rotation, and that co-contraction ratio will explain core kinematics.

## Methods

### *Participants*

The sample size was estimated based upon a power analysis performed using a power of 80% and the effect sizes reported for trunk kinematics (Glofcheskie & Brown, 2017) and PKAM (Sigward et al., 2015, Sigward et al., 2006b) differences between male and female as well as experts and novice (alpha level=0.05). Also, based on Staynor et al. (2020) and Sigward et al. (2015), 62 trials and 22 participants are necessary to achieve an effect size of 0.3 with a power of 80% with maximum 9 predictors of PKAM in the regression models. Accordingly, at least 12 subjects performing 5 trials would be necessary in each group (120 trials in aggregate) and in line with other studies.

The task-expert group was composed of handball players, as sidestepping maneuver is an element of performance for this population. Karatekas were recruited for the nonexpert group. The rationale for the choice of this second population was the rareness of sidestepping maneuvers but a high-level of trunk control in their practice. Comparing experts and non-experts of the task would allow determining the practice-based behaviors to enhance performance and reduce knee joint loading. Thus, the best trunk and pelvis kinematics could be identified and be set as targets for future training programs. Thirteen male handball players (age: 21.5 ± 2.4 years old; height: 1.81 ± 0.06 m; mass: 74.6 ± 7.9 kg) and 14 male karatekas (age: 26.9 ± 8.2 years old; height: 1.77 ± 0.06 m; mass: 72.1 ± 6.3 kg) were included in the study. All participants had at least 10 years of experience in their respective sport. Handball players were playing at the nonprofessional national level and karatekas were black belt. Participants were wearing their own inside sport shoes. No participant had reported an ACL injury or any lower limb or trunk injury during the past 2 months before the experimentation. Prior to testing, all participants were informed about possible risks and gave written informed consent. All procedures were approved by the ethics committee Sud Mediterranee III (approval reference 2018.07.03 bis) and conformed to requirements of the Declaration of Helsinki.

*Protocol*

Participants were asked to perform three different changes of direction tasks on a force plate in a randomized order, including a crossover to -20° to the left, a straightforward deceleration and a sidestepping maneuver to 45° to the right after a dynamic two-step approach. Movement direction was indicated by a light signal triggered at the end of the two-step approach to create an unanticipated change of direction paradigm. Eighteen randomized trials were carried out (6 in each direction) with a 1-minute rest between trials. A trial was defined as successful if the left foot fully hit the force plate and the change of direction was conducted towards the light stimulus directly on the first step after force plate contact.

Moreover, in order to normalize electromyography (EMG) recordings, participants performed a maximal broad jump.

*Material*

The visual light stimulus was triggered automatically via Optojump ™ (Microgate, Bolzano, Italy) on the last step before the force plate contact (Bertec Corp, Columbus, Ohio). Force plate sampling frequency was of 1000 Hz. Light stimulus randomization was computed on Labview (version 15.0.1f1).

Three-dimensional trunk, pelvis and left-leg kinematics were recorded using reflective skin markers (⌀14 mm) placed on anatomical landmarks with self-adhesive tape. Specifically, markers were placed on the upper sternum and the xiphoid process, the C7 vertebra, the anterior superior iliac spines, mid posterior superior iliac spines, the great trochanter, medial and lateral epicondyle of the knee, at the midway of the peroneus, the medial and lateral malleolus, the first and fifth metatarsal, and calcaneus. This marker placement was adapted from the Lyon Whole Body Model (Dumas et al., 2007). Markers were captured with a 6-camera motion analysis system (SIMI Reality Motion Systems, Germany) with a sampling frequency of 100 Hz.

Surface electromyography recordings of core and hip muscles, sampled at 1000 Hz and synchronized with the motion analysis system, were obtained from the rectus abdominis (RAB), the external oblique (EOB), the erector spinae (ESP), the gluteus maximus (GMax) and the gluteus medius (GMed) of the right and left sides. The skin was first shaved, sanded and cleaned with alcohol to reduce impedance, then wireless surface electrodes (Trigno™, Delsys, Natick, MA, USA) were attached parallel to the muscle-fiber orientation, following SENIAM guidelines for sensor locations.

*Data analysis*

All data were only recorded during sidestepping maneuvers to 45°. Marker trajectory and ground reaction force signals were both filtered with a low-pass Butterworth filter (4th order, 15 Hz cutoff frequency) prior to calculating external joint moments with a standard inverse dynamics approach (custom development realized by Moveck Solution Inc.). Three-dimension trunk and pelvis kinematics were defined relative to the global coordinate system. Core kinematics was positive in the direction of the cut, i.e. a rotation towards the right in transverse plane, an anterior tilt in sagittal plane, a lean towards the right in frontal plane (medial lean). A lean towards the left in the frontal plane is therefore defined as a lateral lean. Kinematics and kinetics of left leg were analyzed to determine knee joint loading, with a focus on peak knee abduction moment (PKAM), as this latter parameter is associated with knee injury (Hewett et al., 2005). All EMG data were band-pass filtered (10 Hz-500 Hz). Muscle activation of the broad jump was then low-pass filtered at 30 Hz prior to determining a reference peak value during the jump. Heart muscle electrical activity was filtered with a Butterworth high-pass filter at 30 Hz on the RAB and the EOB. Root mean square (RMS) values were determined during the preactivation phase (100 ms prior to the initial contact with the force plate) and during the weight acceptance (WA) phase, which represented the time between IC and the first trough in the vertical force. The activation of the different muscles was normalized to their peak filtered value recorded during the maximal broad jump. The simultaneous action of agonist and antagonist muscles was analyzed via directed co-contraction ratios (DCCR) during the preactivation phase and WA (Donnelly et al., 2015). DCCR were defined according to the anatomical function of the muscles with respect to motion during the change of direction. Therefore, a positive ratio in trunk flexion/extension would indicate a higher co-contraction of RAB and EOB (agonists) than ESP (antagonists). In the frontal plane, right EOB and ESP were defined as agonists and left EOB and ESP were antagonists. According to the change of direction to the right, left EOB was defined as the agonist and right EOB was the antagonist with respect to trunk rotation. Although fewer muscles were recorded at the pelvic level, the anterior/posterior pelvic tilt could also be analyzed using directed co-contraction ratio, where a positive ratio would indicate a higher co-contraction of ESP (agonists) than RAB, EOB and GMax (antagonists). DCCR values were between -1 (towards antagonists) and 1 (towards agonists), while a zero would indicate equal activation of agonist and antagonist muscle groups.

*Statistics*

All variables were averaged across the six sidestepping trials for each participant. Results are presented as group mean (standard deviation, SD). Normal distribution and variance homogeneity were respectively verified via Lilliefors' and Levene's test. The influence of the sidestepping expertise (Handball vs. Karate) on the dependent variables during changes of direction was analyzed using independent Student's t-tests. The level of significance was set at $p < 0.05$. Moreover, for each significant variable, effect sizes (ES) were calculated with Cohen's d (0.2, 0.5 and 0.8 correspond to small, medium and large effects) and confidence intervals (CI) of 95% were reported.

A total of 162 trials were used to perform linear mixed model analyses. Such models take account of within-participant effect, allowing an analysis with individual trials as data points. First, two linear mixed models were used to determine the association between PKAM

(dependent variable) and independent variables. At IC, the predictors were 3D core kinematics (model no.1) and during WA, 3D GRF were added to 3D core kinematics (model no.2). Sidestepping expertise was included as fixed effect covariates and subject effect as a random effect. Backwards elimination of independent variables (fixed effect) was performed by sequentially removing nonsignificant predictors until all predictors were significant. Then, kinematic predictors of PKAM were defined as dependent variables of linear mixed models at IC and during WA. Trunk DCCR values were the independent variables for trunk kinematics and RMS of the gluteal muscles and pelvis DCCR were the predictors for pelvis kinematics, during the preactivation phase(model no.1) and WA (model no.2). The same backwards elimination procedure was processed.

Standardized effects of each predictor were computed by multiplying the predictor standard deviation with its estimate. Estimate values were coefficients of independent variables in a linear mixed model. Standardized effects were expressed with respect to the mean of the dependent variable, i.e., the one SD change of the dependent variable when all other predictors were set to their average value (Staynor et al., 2020).

## Results

### *Kinetics and kinematics*

At IC, none of the trunk kinematical variables differed between handball players and karatekas (Table 1). Pelvis rotation to the right was greater for karatekas than handball players ($p<0.01$). Its mean difference was -5.9° (CI = [−10.1; −1.7]; ES = 1.11).

|  | Handball | Karate | Pooled subjects |
| --- | --- | --- | --- |
| **Trunk medial lean** | -3.8 (2.6) | -3.1 (3.4) | -3.4 (4.4) |
| **Trunk axial rotation** | -4.3 (6.3) | -6.9 (5.5) | -5.6 (7.0) |
| **Trunk anterior tilt** | 12.9 (5.2) | 16.0 (6.0) | 14.5 (6.2) |
| **Pelvis medial lean** | 1.1 (2.6) | 1.7 (2.9) | 1.4 (3.6) |
| **Pelvis axial rotation** | **4.9 (6.0)** | **10.8 (4.5)\*** | 7.9 (6.6) |
| **Pelvis anterior tilt** | 10.0 (4.1) | 11.0 (3.8) | 10.5 (4.4) |
| **Knee abduction angle** | 2.4 (1.4) | 0.21 (4.1) | 1.3 (3.4) |

Table 1. Comparison of mean (SD) kinematic variables

During WA (Table 2), PKAM was significantly greater for handball players than for karatekas (p = 0.04), as well as peak medial and vertical GRF (p = 0.01). The mean difference of PKAM was 0.218 Nm/kg-bwt (CI = [0.008; 0.427]; ES = 0.84). Finally, the mean difference of peak medial GRF was 2.2 N/kg-bwt (CI = [0.58; 3.80]; ES = 1.03) and 3.6 N/kg-bwt (CI = [0.84; 6.34]; ES = 1.01) for peak vertical GRF.

|  | Handball | Karate | Pooled subjects |
| --- | --- | --- | --- |
| **Trunk medial lean** | -12.1 (6.5) | -11.7 (4.6) | -12.0 (6.1) |
| **Trunk axial rotation** | -2.4 (8.2) | -4.8 (8.6) | -3.6 (9.3) |
| **Trunk anterior tilt** | 18.5 (5.9) | 20.4 (6.4) | 19.6 (6.7) |
| **Pelvis medial lean** | 0.2 (3.4) | 1.2 (3.1) | 0.7 (4.1) |
| **Pelvis axial rotation** | 3.6 (6.7) | 7.7 (5.5) | 5.7 (7.1) |
| **Pelvis anterior tilt** | 7.6 (4.4) | 8.6 (4.8) | 8.1 (5.0) |
| **Knee abduction angle** | 2.7 (1.7) | 1.2 (4.3) | 1.9 (3.7) |
| **Peak knee abduction moment** | **0.539 (0.302)** | **0.321 (0.209)*** | 0.425 (0.497) |
| **Peak medial GRF** | **13.8 (2.2)** | **11.6 (1.8)*** | 12.6 (2.7) |
| **Peak posterior GRF** | 6.0 (2.3) | 4.7 (2.0) | 5.3 (2.5) |
| **Peak vertical GRF** | **27.8 (4.0)** | **24.2 (2.7)*** | 26.0 (4.5) |

Table 2. Comparison of mean (SD) kinematic variables (°), peak knee abduction moment (Nm/kg-bwt) and GRF (N/kg-bwt) between handball players and karatekas during weight acceptance phase. *: p<0.05

*Neuromuscular control*

Neither DCCR values nor glutei RMS values were significantly different between handball players (HB) and karatekas (K) (Figure 1). During the preactivation phase, pelvis DCCR supported pelvis posterior tilt (HB: −0.15 ± 0.49; K: −0.22 ± 0.31) and during WA phase, pelvis DCCR remained negative (HB: −0.26 ± 0.37; K: −0.03 ± 0.36).

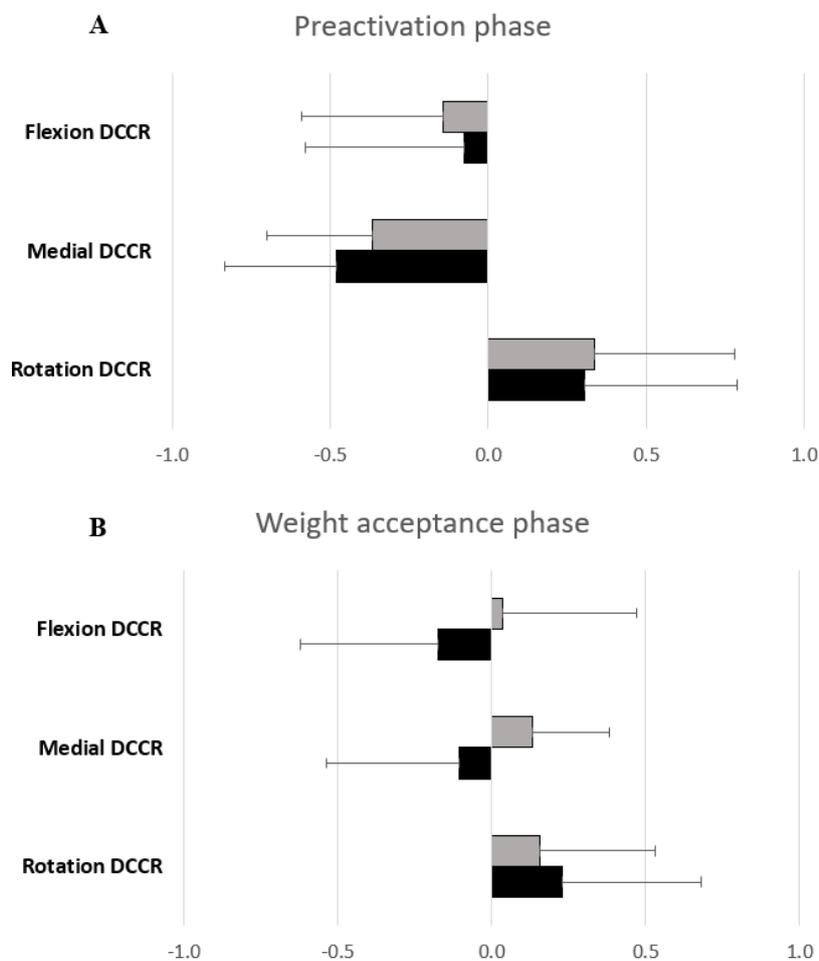

Figure 1. Directed Co-Contraction Ratio (DCCR) during the preactivation phase (A) and the weight acceptance (WA) phase (B) for karatekas (grey bars) and handball players (black bars). Positive values indicate a co-contraction ratio towards trunk flexion, trunk medial lean and trunk right rotation.

*Knee joint loading and core kinematics predictors*

Predictors of PKAM and predictors of significant kinematic variables are reported in Table 3. In the first model, at IC, PKAM was negatively associated with trunk medial lean and increased with greater trunk axial rotation (towards the new direction) and pelvis anterior tilt. In the second model, during WA, PKAM increased with larger trunk axial rotation (towards the new direction) and peak posterior GRF, but was reduced when pelvis medial lean increased. Moreover, the different kinematical variables could be predicted by trunk and pelvis neuromuscular control.

|  | Predicted variable | Independent variable | Estimate | CI | One SD change (%) |
|---|---|---|---|---|---|
| IC | PKAM (model no. 1) | Trunk medial lean | -33.9 | [-51.9; -15.9] | 35 |
|  |  | Trunk axial rotation | 13.1 | [1.6; 24.6] | 22 |
|  |  | Pelvis anterior tilt | 20.4 | [2.3; 38.5] | 21 |
|  | Trunk medial lean | Medial DCCR | 3.11 | [1.38; 4.85] | 35 |
|  | Trunk axial rotation | Flexion DCCR | -3.54 | [-5.56; -1.51] | 33 |
|  | Pelvis anterior tilt | Pelvis DCCR | 2.93 | [1.49; 4.38] | 13 |
| WA | PKAM (model no. 2) | Trunk axial rotation | 11.8 | [3.0; 20.6] | 26 |
|  |  | Peak posterior GRF | 40.9 | [10.0; 71.7] | 24 |
|  |  | Pelvis medial lean | -23.2 | [-43.7; -2.7] | 22 |
|  | Trunk axial rotation | Flexion DCCR | -7.00 | [-9.64; -4.36] | 98 |
|  | Pelvis medial lean | Left Gluteus Medius | 0.030 | [0.011; 0.049] | 164 |
|  |  | Right Gluteus Medius | -0.026 | [-0.044; -0.007] | 149 |
|  |  | Right Gluteus Maximus | -0.081 | [-0.156; -0.006] | 95 |

Table 3. Kinematic and kinetic variables at initial contact (IC) and during weight acceptance (WA) predicting peak knee abduction moment (PKAM). EMG variables (DCCR or RMS) predicting core kinematics significant PKAM predictors. CI: Confidence interval; DCCR: directed co-contraction ratio. Gluteus Medius and Gluteus Maximus independent variables refer to their RMS activity.

**Discussion and Implications**

The present study's objectives were to determine the differences of core stability and knee joint loading between experts and nonexperts of sidestepping maneuvers and the influence of core stability on a knee joint injury predictor. The main findings were: 1) that there was no evidence of major biomechanical differences or injury risk during a sidestepping maneuver between handball players and karateka, despite a higher PKAM and lower pelvis rotation towards the new movement direction for handball players; 2) that trunk and pelvis kinematics predicted

PKAM; and 3) that directed co-contraction ratio could help in the understanding of trunk and pelvis kinematics.

*Core stability and knee joint loading differences between experts and nonexperts*

Despite significantly higher PKAM for handball players, which validated our hypothesis, it could not be considered as a higher risk of ACL injury, as its mean value remained relatively low and was in line with the literature (Sigward et al., 2015). Moreover, Hewett et al. (2005) reported PKAM values on ACL injured females of around 0.73 Nm/kg-bwt during drop jump while uninjured females were at 0.31 Nm/kg-bwt. Furthermore, using a biomechanical model, Lin et al. (2009) estimated ACL injury risk for males to increase significantly with a PKAM of around 1.25 Nm/kg-bwt. Thus, handball players maintained knee joint loading at a relatively low level, unlikely to be associated with ACL injury. Otherwise, lower PKAM values for karatekas might find an explanation in their absence of expertise in the sidestepping task, yielding more challenging changes of direction execution for this population. In order to keep their joints safe while experiencing a new task, karatekas might have adopted some protective behavior. This would be in line with Sigward and Powers (2006b), who reported higher PKAM values for experts than nonexperts in the same sport. One potential adaptation could be the pelvis orientation towards the new direction already at IC. As the pelvis is central for core stability (Houck et al., 2006; Weltin et al., 2016), it could be suggested that better orienteering of the pelvis in the new sidestepping direction might stabilize the whole body and be a protective strategy for the knee. However, the sidestepping task's objective is to eliminate defender in handball. An obvious rotation toward the direction would give early information to the opponent, explaining why handball players presented a more neutral position. Greater lateral and vertical GRF were observed in handball players, whereas knee joint loading was kept at a safe level, suggesting a more performance-orientated execution by handball players, which is in line with the potential more aggressive behavior hypothesized by Sigward and Powers (2006b). Despite pelvic kinematics changes, no trunk motion difference was observed. Thus, contrary to our hypothesis, handball players did not better orientate their trunk in the targeted direction, despite differences in trunk stability among athletes with different disciplines and levels of practice reported in the literature (Barbado et al., 2016; Gloscheskie & Brown, 2017). However, these studies revealed that athletic background doesn't necessarily impact every component of the core stability (e.g. core proprioception evaluated during seated postural control; Gloscheskie & Brown, 2017). Therefore, our results show that handball players and karatekas might have quite similar core stability and thus karatekas could transfer their ability to control their trunk to this new sidestepping task. In addition, the realization of unanticipated sidestepping might imply too high constraints, which subjects would handle almost in the same manner whatever the level of expertise or core stability.

*Kinematic predictors of injury risk*

In the frontal plane, this study has confirmed previous observations of an increase of trunk lateral lean at IC increasing PKAM (Dempsey et al., 2012; Jamison et al., 2012; Jones et al., 2015; Kristianslund et al., 2014; Mornieux et al., 2014; Staynor et al., 2020). Moreover, pelvis lateral lean during WA phase increased PKAM. Trunk lateral lean had the highest one SD percentage of change meaning that this variable explained more variation of PKAM than any others in the model. Thus, training interventions seeking to reduce ACL injury risk should primarily target trunk control in the frontal plane.

In the transverse plane, larger rotation towards the new direction was associated with larger PKAM. If trunk rotation towards the stance limb seems to be a predictor of PKAM during landings (Critchley et al., 2020; Dempsey et al., 2012) and planned sidestepping maneuvers (Frank et al., 2013), our results are in line with previous studies on unplanned sidestepping maneuvers (Kristianslund et al., 2014; Staynor et al., 2020). Therefore, trunk rotation during changes of direction should be taken into account with caution. While increased trunk rotation towards the new movement direction could be seen as a strategy to reorient the whole body and improve performance (Marshall et al., 2014), we found that this could also be associated to some extend with increased knee joint injury risk.

In the sagittal plane, only the pelvis anterior tilt at IC appeared to predict PKAM. From an anatomical point of view, pelvis anterior tilt flexes and rotates the hip internally, which changes gluteal muscles moment arms and weakens the hamstrings, therefore possibly impacting PKAM (Alentorn-Geli et al., 2009). Overall, our results confirm the importance of an adapted core stability to limit ACL injury risk (Larwa et al., 2021).

*Neuromuscular predictors of core kinematics and knee injury risk*

As no specific trunk kinematics differed between our two populations, seeing no difference in directed co-contraction ratio was not surprising. However, trunk and pelvis neuromuscular control predicted core stability. DCCR in the frontal plane explained trunk lateral lean at IC, suggesting that higher contractions towards the stance leg during the flight-phase supported trunk lateral lean. Later on, during WA, lateral pelvis lean was predicted by both GMed and the right GMax. According to Neumann (2010), this might underline the role of the left GMed (stance leg) as being associated with the pelvis contralateral drop, while right Gmed and GMax muscles probably abducted the contralateral lower limb. Surprisingly, DCCR in rotation did not predict trunk rotation, while DCCR in flexion did. DCCR in rotation was computed, based on only external oblique muscles, whose 3D actions might have blurred their impact on trunk rotation.

Pelvis anterior tilt at IC was predicted by anterior-posterior DCCR, meaning that core muscles acted to limit excessive pelvis anterior tilt. However, as DCCR explained only 13% of the pelvis tilt, the role of each muscle integrated in this ratio might not be fully understood. For instance, gluteus maximus probably also acted to control hip flexion as antagonists (Alentorn-Geli et al., 2009) and maybe had a larger role in force absorption than tilting the pelvis (Powers, 2010).

*Limitations*

The dynamic two-step approach used in the present study was similar to handball practice but differed from most of the setups used in previous changes of direction studies with a classical run-up at a given approach speed. Nevertheless, PKAM values and trunk kinematics were in the range of the results found in the literature with a classic 4 $m.s^{-1}$ running speed (Weir et al., 2019; Mornieux et al., 2021) or a 3 steps approach (Jamison et al., 2012). Also, the sampling frequency used for kinematics recording (i.e. 100 Hz) was lower than the 200 Hz typically reported in the literature. However, given the maximal frequency of the kinematic signals during changes of direction (up to 20 Hz), the present sampling frequency should be enough to gather the full information of the kinematic signals. Finally, trunk neuromuscular control could not be fully evaluated thanks to surface EMG, and future research might want to add an evaluation of deeper muscle activity.

**Conclusion**

Nonexperts in changes of direction reduced ground reaction forces and knee joint abduction moment by potentially rotating their pelvis more towards the new movement direction during sidestepping. However, the higher level of knee joint loading during the sidestepping task in experts might not be seen as a higher risk of ACL injury. Moreover, this study confirms that core stability (especially the trunk in the frontal plane) predicts knee joint loading and therefore possibly ACL injury risk. Finally, our results might guide neuromuscular prevention programs to focus on trunk lateral flexors' eccentric action in order to limit trunk lateral lean and on gluteal muscles' ability to stabilize pelvic lateral lean. This research could also influence interventions on kinematic changes of athletes' techniques while sidestepping. Further research could explore the balance between performance and injury during changes of direction tasks.

**Acknowledgments**

The authors would like to thank all the participants of this study and Kristin L. Sainani for the statistical help.

**Disclosure statement**

No potential conflict of interest was reported by the authors.

**Appendices**

Figure 2: Typical filtered signals (mV) for rectus abdominis (RAB), external oblique (EOB), erector spinae (ESP) left and right muscles and ground vertical force (Fz). The first lightly shaded area represents the pre-activity (100ms) prior to the initial contact (IC) and the second shaded area is the weight acceptance phase (WA). The vertical dashed line delimits the end of the foot-contact during the sidestepping task.

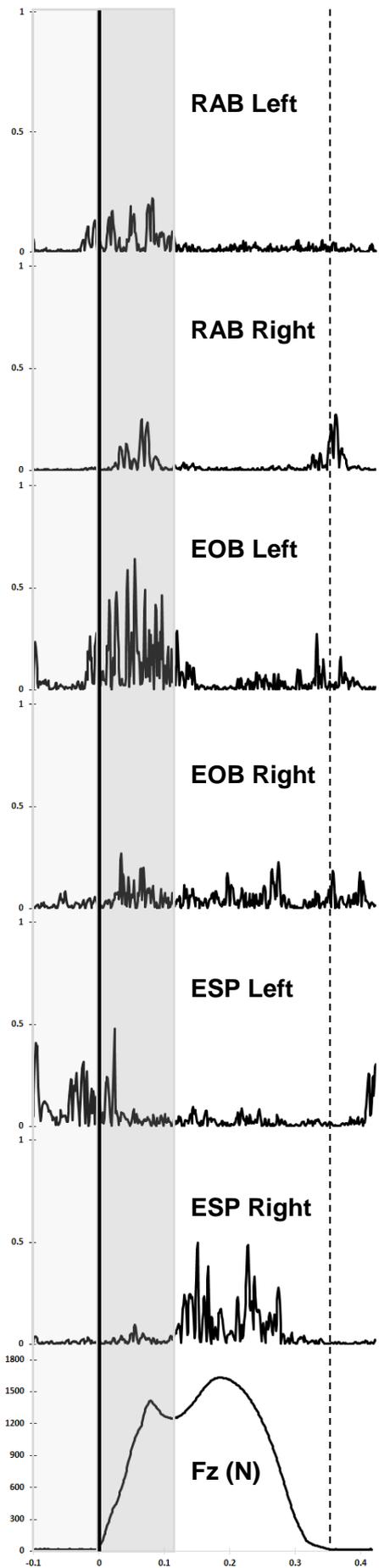